\documentstyle[prl,aps,multicol,psfig]{revtex}
\begin{document}
\draft
\newcommand{\be}{\begin{equation}}
\newcommand{\ee}{\end{equation}}
\newcommand{\ba}{\begin{eqnarray}}
\newcommand{\ea}{\end{eqnarray}}

\title{Edge of a Half-Filled Landau Level}
\author{S.-R. Eric Yang$^{1,2}$ and J. H. Han$^{2}$}
\address{Department of Physics, Korea University, Seoul 136-701,
Korea$^{1}$
\\ and \\
Asia Pacific Center for Theoretical Physics, Seoul, Korea$^{2}$}
\maketitle
\draft
\begin{abstract}
We have investigated the electron occupation number of
the edge of a quantum Hall (QH) droplet at $\nu=1/2$ using
exact diagonalization technique and composite fermion
trial wavefunction.
We find that the electron occupation numbers near the edge  
obey a scaling behavior.  The scaling result indicates the existence of
a well-defined edge corresponding to the radius of a
compact droplet of uniform filling factor 1/2. We find that 
the occupation number beyond this edge point
is substantial, which is qualitatively different from
the case of odd-denominator QH states. We relate these features 
to the different ways in which composite fermions occupy Landau levels for
odd and even denominator states.
\end{abstract}
\thispagestyle{empty}
\begin{multicols}{2}

The fermion Chern-Simons theory of the half-filled Landau level has provided 
a simple and useful picture of the bulk
state\cite{halperin,rezayi,recent}.
According to this theory effective  
composite particles move in the absence of external magnetic field and
form a compressible metallic state.
There is a strong current interest in the edge properties of this novel
liquid\cite{brey,chklovskii,chang,conti,han,shytov,khv}.
However, unlike odd-denominator  QH liquids, it is difficult to
construct a gauge invariant chiral Luttinger theory\cite{wen,kane}
at $\nu=1/2$ since 
the longitudinal conductivity is non-zero.
Experimentally the electron tunneling I-V characteristic at the edge is
found to behave as
$I(V)\sim V^g$ with a $\nu$-dependence of the tunneling exponent 
$g\approx \nu^{-1}$\cite{chang}.
Various theories of the one-particle spectral density in the edge state of
compressible fractions exist\cite{conti,han,shytov,khv}. 
There are some discrepancies in the prediction of the tunneling exponent 
among these theories. 

Microscopically the fundamental properties of the edge are not
well known in the case of compressible fractions such as 1/2. 
Also it is not clear 
how their edge shapes compare with those of  incompressible
QH states. In the light of this,
the electron occupation number $n_k$ serves as a useful quantity to
investigate since 
it can be compared directly with the well-known result for incompressible
QH states at the filling factor $\nu=1/M$: $n_k\propto
(k-k_{ed})^{M-1}$, where
the wavevector corresponding to the position of the edge is
$k_{ed}=Mk_F$ 
and $k_F$ is the Fermi wavevector\cite{mitra,yang2}.

An accurate calculation of the values of $n_k$ near the edge
requires a correct description of 
strong electron correlations using non-perturbative methods.
In this paper we have investigated the properties of the compressible
$\nu=1/2$ edge
using finite size exact diagonalization (ED) and finite-size scaling
analysis,
followed by a variational composite fermion  wavefunction
approach\cite{jain}.
When a symmetric gauge is used the single particle states at
large magnetic fields are characterized by the Landau level index $n$ and
angular momentum number $m$.  For given $n$ the possible values of angular
momentum are $m=-n,-n+1,...,0,1,...$.  
We have computed the ground state distribution of the occupation number
$n_m$
on a disk
for up to eleven particles ($m$ is related to the 
wavevector $k$ through $k=2\pi m/R_m$, where
$R_m=\sqrt{2(m+1)}$ is the mean radius of the angular momentum state
$m$.) We set the magnetic length to one. We find: 
(a) The occupation number $n_m$ of the edge 
satisfies a scaling relation so that the values of a big
system may be extracted from those of relatively small systems.
(b) The occupation number $n_m$ is non-zero
even for $m$ greater than $m_{ed}=2N-1$.
This is in sharp contrast to the $N$-particle incompressible states where
$n_m$ is non-zero only up to $m_{ed}=M(N-1)$. A variational wavefunction
study
also leads to the similar conclusion.
(c) Numerically  
we find that $n_m$ changes abruptly near $m=m_{eff}$,
where $n_m$ is very small beyond $m=m_{eff}$. 
(d) In the  composite fermion picture, {\it some} of the composite
fermions occupy 
higher (effective) Landau levels of composite fermions 
while the fractional Laughlin state is obtained when {\it all} of
the composite fermions lie in the lowest Landau level. The tail region of $n_m$
reflects this occupation behavior of composite fermions.

In the cleaved-edge construction used
by Grayson {\it et al.} in their tunneling experiments, the edge
potential
is believed to be very sharp\cite{chang}.
The basic physics of this
situation may be modeled by fixing the total angular momentum of the
droplet
since for a certain range of the confinement strength the groundstate
is given by the half-filled state with the total angular momentum
$M_{TOT}=N(N-1)$ \cite{yang1}.
The maximum possible value of occupied
angular momentum state may be determined as follows: First note that
arbitrary states in the lowest Landau level may be written in the form 
\be
 S[\{z_i\}]\prod_{i<j}(z_{i}-z_{j})
           \exp(-\sum_{i}|z_{i}|^2 /4),
\ee
where $S[\{z_i\}]$ indicates an arbitrary {\it symmetric} 
polynomial\cite{trugman}.
Since the total angular momentum of a half-filled $N$-particle state is
required to be $N(N-1)$ and the Jastrow factor $\prod_{i<j}(z_{i}-z_{j})$
carries the angular momentum $N(N-1)/2$, the remaining $N(N-1)/2$ units of
angular momenta are carried by the symmetric polynomial. Therefore the
angular momentum for a single particle state may be as large as the degree
of the symmetric polynomial plus those from the Jastrow factor, which
equals $m_c=(N-1)+N(N-1)/2=(N+2)(N-1)/2$.

Figure 1 displays the ED results for $n_m$ at various values of $N$.
The oscillatory nature of the occupation numbers
in the droplet had also been found for the incompressible
fractions\cite{mitra}. 
Explicit analytical result is available for $N=3$ since the
exact groundstate wavefunction is known for $(N,M_{TOT})=(3,6)$
\cite{laughlin}.
The result of the ED agrees with high accuracy with this analytical result. 
The shape of the occupation number is rather different from that of 
$1/3$ state\cite{mitra}.  
A striking feature of the ED result is that
the occupation number exhibits a tail above $k_{ed}$ 
(to be defined shortly) at $\nu=1/2$. This property clearly
departs from that of the incompressible edge where, for a sharp confining
potential, the maximum occupation is fixed at $M(N-1)$. 
A sudden drop of
occupation number occurs at $k$ slightly greater than $k_{ed}$ is
observed
for all our ED calculations, as illustrated in Tables I and II. For $N=7$
to $10$ particles,
the drop in the occupation occurs when $m>m_{eff}$ with
$m_{eff}=15,17,20$,
and $21$ respectively.  The ED results show that the
effective cut-off radius $m_{eff}$ is greater than $m_{ed}$, but still
much
smaller than the theoretical maximum value of $m_c$. 

Figure 2 displays a scaling function that $n_m$ of the edge obeys
approximately
\begin{eqnarray} n_m=f((k-k_{ed})R_{ed}). 
\label{eq:scaling}
\end{eqnarray}
We have defined
$m_{ed}=2N-1, R_{ed}=\sqrt{2(m_{ed} +1)}$, and $k_{ed}=2\pi m_{ed}/R_{ed}$
respectively. The choice of our length scale $R_{ed}$ is motivated by the
fact that it is  the radius of the compact droplet whose average
filling factor is 1/2.
As we move to the edge some kind of scaling behavior is
already apparent for the system sizes we have investigated.  Since
$k_{ed}R_{ed}=2\pi (2N-1)$, the starting $x$-axis values in the figure 
are displaced
horizontally by a fixed amount $4\pi$ for each increment of the particle
number.  While the peak positions of the occupation do not occur at the
same point in the scaling variable we have chosen, the downward slide to
zero occupation obviously follows a single curve. Finite-size
corrections are visible in the increasing edge slope and the dwindling
peak value for larger $N$. We expect that these corrections will ultimately
vanish when we go to a large enough system.
The occupation 
values seem to cross at $k=k_{ed}$. 
We have verified that the same
data plotted as $f(k/k_{ed})$ also exhibit crossing at 
$k/k_{ed}\approx 1$. 
The presence of $k_{ed}$ signifies the existence of a well-defined edge 
in the compresible states. 
The difference of
$R_{ed}$ and the cut-off radius where the occupation nearly
vanishes,
$W=R_{eff}-R_{ed}$ with $R_{eff}=\sqrt{2(m_{eff} +1)}$, is zero for the
incompressible edges.  
From our numerical results we find that $W$ is nonzero in the thermodynamic
limit for compressible edges.


Several key points of the ED results may be understood from trial variational
wavefunction approach. These are: 
(a) $m_{eff}\not=m_c,m_e$ and grows with $N$, and (b)
there is a sudden drop in $n_m$ for $m>m_{eff}$.
A good trial variational wavefunction for states 
close to half-filling 
is obtained by attaching two vortices to each electron:
\be
\Psi=P_{LLL}[\prod_{i<j}(z_i-z_j)^2\phi_{M^*}].
\label{eq:CFwavef}
\ee
The Slater determinant state $\phi_{M^*}$ consists of $N$
Landau level states $(n_i, m_i)$. For the half-filled state where the
residual flux is zero, Rezayi and
Read proposed to use the Slater determinant of free fermions for
$\phi_{M^*}$\cite{rezayi}. Recently Jain and Kamilla (JK) proposed an
alternative way of
writing the projected wavefunction in a symmetric gauge\cite{jain}. 
In our geometry it is more convenient to use JK's  wavefunction.
While JK's analysis  focuses on 
the {\it bulk properties} of their variational wavefunction,
here we focus on the edge properties of such states. As
will become clear below, the variational state proves to be an excellent
approximation to our ED calculation in many respects.

Since the total angular momentum of the groundstate at half-filling is
$N(N-1)$, we obtain $M^*=\sum_{i=1}^{N}m_{i}=0$. 
For $N=7,6,5$, JK have chosen the projected groundstate
with $M^*=0$  as the compact states
[4,1,1,1], [3,2,1], and [3,1,1]\cite{jain}. 
The agreement in the ground state energy computed in terms of JK's
composite fermion wavefunction with the exact result is exceedingly good.
Encouraged by this, we have computed the {\it occupation
number} using JK's wavefunction and compared them with the exact
results. We list the occupation numbers for $N=5$ in Table II.
For $N\ge 6$, the number of terms in the Slater determinant is too large
to be handled exactly.
As one can see in Table II, the agreement of numbers is 
excellent except at either ends of the distribution, where the relative
difference is about 10 \%. 
The sum of the occupation numbers in both cases is 5 to high
accuracy. 
One can infer that JK's function is built out of a
somewhat restricted set of states compared with the true ground state from
the fact that small occupation probabilities at $m>m_{eff}$
are replaced by zeroes in JK's wavefunction. For generic $N$, the true
ground state is likely to be a sum of several states of a given
total angular momentum in which JK's state makes the most dominant
contribution. 
 
To estimate the last occupied state for general $N$, we need to know how to
write down JK's wavefunction for arbitrary number of particles.
For  $N=2q$ or $2q+1$, we choose the trial wavefunctions as
the following compact states
\ba
 & [q,2,\overbrace{1,1,..,1}^{q-2}] \nonumber \\
 & [q+1,\underbrace{1,1,..,1}_{q}],
\label{eq:JainState}
\ea
where $q$ is a positive integer. 
When other states are available with
the same $M_{TOT}$, we must choose the one with the lowest energy.
According to JK's analysis, the state which creates the least number of
defects also has the lowest energy. By maximizing $N_0$, we can minimize
the creation of the defects. Such consideration leads to the above
states\cite{note}.
We will
show that the above choice do correctly lead to some of the
features observed numerically.

The total angular momentum of the first state is $M_{TOT}=2q(2q-1)$ and for the
second state it is $M_{TOT}=2q(2q+1)$. Using the states described in Eq.\
(\ref{eq:JainState}), one can show that the last occupied single particle
states occur at $m=5q-3$ and $5q$ for even and odd $N$ respectively.  This
gives $m_{eff}=15,17,20$, and $22$ for $N=7-10$ in good comparison with the ED
results $m_{eff}=15,17,20,$ and $21$. Using the values of $m_{eff}=5q-3$ and
$5q$ we can estimate $W \propto \sqrt{N}$.

Our trial state, Eq.\ (\ref{eq:JainState}), shows that about half the 
composite fermions occupy the lowest Landau level, while the higher Landau
levels are singly occupied by the other half. This is in contrast to the
Laughlin states which, in JK's notation, is given by $N_0=N$. That is, all
of the composite fermions occupy the lowest Landau level to form an
incompresible Laughlin state. Even in the case when Eq.\ (\ref{eq:JainState})
is not the groundstate, other possible states will also have a
distribution of
composite fermions in higher Landau levels.
We believe such differences in the
occupation behavior of composite fermions explain the observed difference
of {\it electron occupation} for compressible/incompressible states.

In conclusion, our results indicate that 
the behavior of the edge at $\nu=1/2$
is substantially different from that of the incompressible case.  
We find also that a sizable fraction of composite fermions  
occupy the higher Landau levels,
which is  reflected in the existence of a tail in $n_m$.
It would be useful to calculate analytically the scaling
function $f(x)$. Doing so would first require the construction of a proper
effective theory of the edge which, like Wen's theory\cite{wen}, will predict
both the dynamic behavior (Green's function) and the ground-state behavior like
the occupation number. It is hoped that our results may be useful
in understanding such an effective low energy theory of the edges of
compressible states.

This work has been supported by  the KOSEF
under grant 961-0207-040-2 and the Ministry of Education under grant
BSRI-96-2444.

\end{multicols}

\newpage
\begin{table}
\caption{Occupation number for $N=$7 obtained from exact diagonalization.
Note that it drops suddenly at $m=16$.}
\begin{tabular}{cccccccccc}
 $m$ & 0 & 1 & 2 & 3 & 4 & 5 & 6 & 7 & 8 \\
 $n_m$ & .3648 & .3253 & .4050 & .6037 & .7353
 & .7963 &.8265 & .6974 & .6221\\ \hline
$m$  & 9 & 10 & 11 & 12 & 13 & 14 & 15 & 16 & 17 \\
$n_m$ &  .5359 & .3526 & .2714 & .2011 & .1470 & .0821  &
  .0325  & 7.85$\times 10^{-4}$ & 2.71$\times 10^{-4}$

\end{tabular}
\end{table}

\begin{table}
\caption{Comparison of occupation numbers for $N=5$; ED vs. 
JK's trial wavefunction.
Note that the ED result at $m=11$ is $6.0\times 10^{-5}$ while that of the
trial wavefunction is $0$.}
\begin{tabular}{ccccccc}
 $m$ & 0 & 1 & 2 & 3 & 4 & 5 \\
ED & .1805 & .3671 & .7375 & .8442 & .9012 & .8035 \\
JK & .2075 & .3750 & .7285 & .8298 & .8910 & .7820 \\ \hline
 $m$ & 6  & 7 & 8 & 9 & 10 & 11 \\
ED & .6208 & .2549 & .1562 & .0967 & .0373 & 6.0$\times 10^{-5}$ \\
JK & .6076 & .2608 & .1692 & .1069 & .0419 & 0
\end{tabular}

\end{table}

\begin{figure}
\caption{Occupation numbers for $N=8 (\circ),9 (\Box),10
(\Diamond),11 (\triangle)$. Extremely small occupation
numbers at large $m>m_{eff}$ are not shown.}
\end{figure}
\begin{figure}
\caption{Occupation numbers of Fig.\ 1 in a scaling form
$n_m=f((k-k_{ed})R_{ed})$. The quantities $k_{ed}$ and $R_{ed}$ 
are defined in the text.}
\end{figure}


\begin{references} 
\bibitem{halperin}B. I. Halperin, P. A. Lee, and N.
Read, Phys. Rev.  B {\bf 47}, 7312 (1993). For a recent survey of the
subject, see  {\it Perspectives in Quantum Hall Effects}, edited by S. Das
Sarma and A. Pinczuk (John Wiley \& Sons, 1997). 
\bibitem{rezayi}E. Rezayi, and N. Read, Phys. Rev. Lett. {\bf 72},
900 (1994).
\bibitem{recent}N. Read, Semi. Cond. Sci. Tech. {\bf 9}, 1859 (1994);
R. Shankar and Ganpathy Murthy, cond-mat/9702098; D. H. Lee,
cond-mat/9709233.
\bibitem{brey}L. Brey, Phys. Rev. B {\bf 50}, 11861
(1994).
\bibitem{chklovskii}D. B. Chklovskii, Phys. Rev. B {\bf 51}, 9895 
(1995).
\bibitem{chang} M. Grayson, D. C. Tsui, L. N. Pfeiffer, K.
W. West, and A. M. Chang, To be published.
\bibitem{conti} S. Conti and G. Vignale, Phys. Rev. B {\bf 54}, 14309
(1996); S. Conti and G. Vignale, cond-mat/9709055.
\bibitem{han} J. H. Han and D. J. Thouless, Phys. Rev. B {\bf 55},
1926 (1997); J. H. Han, Phys. Rev. B {\bf 56}, 15806 (1997).
\bibitem{shytov} A. V. Shytov, L. S. Levitov, and B. I. Halperin,
cond-mat/9703246.
\bibitem{khv} D. V. Khveshchenko, cond-mat/9710137.
\bibitem{wen}For a review see X. G. Wen, Int. J. Mod. Phys. B
{\bf 6}, 1711 (1992).
\bibitem{kane} C. L. Kane and M. P. A. Fisher, 
{\it Perspectives in Quantum Hall Effects}, edited by S. Das
Sarma and A. Pinczuk (John Wiley \& Sons, 1997).

\bibitem{mitra} Sami Mitra and A. H. MacDonald, Phys. Rev. B {\bf 48},
2005 (1993).
\bibitem{yang2}E. H. Rezayi and F. D. M. Haldane, Phys. Rev. B
{\bf 50}, 17199 (1993); S.-R. Eric Yang, Sami Mitra, A. H. MacDonald, and
M. P. A. Fisher, J. Korean Phys. Soc. (Proc. Suppl.)
{\bf 49}, S10 (1996).
\bibitem{jain} J. K. Jain, and R. K. Kamilla, cond-mat/9704031;
For a composite fermion
occupying the $n$-th (effective) Landau level and the angular momentum
$m$, the
orbital wavefunction is given by
\be
\eta^{CF}_{n,m}(z)=
N_{n,m}e^{-|z|^2/4}z^{m+n}\partial^n
\prod_{k}(z-z_k).
\ee
The maximum angular momentum quantum number associated with $z$
is $(N-1)+m$. The ground state is chosen as the compact state
$[N_0, N_1, N_2,...]$ with $N_0\geq N_1\geq N_2,...$, where $N_i$ is the
number of composite fermions in the $i$-th (effective) Landau level.

\bibitem{yang1}S.-R. Eric Yang, A. H. MacDonald, and M. D. Johnson,
 Phys. Rev. Lett. {\bf 71}, 3194 (1993).
\bibitem{trugman} S. A. Trugman and S. Kivelson, Phys. Rev. B
{\bf 31}, 5280 (1985).
\bibitem{laughlin}R. B. Laughlin, Phys. Rev. B {\bf 27}, 3383 (1983). 
\bibitem{note}For $N\le 8$, numerical results of JK and ours show that the
states given by Eq.\ (\ref{eq:JainState}) is the correct ground state. 
\end{references}
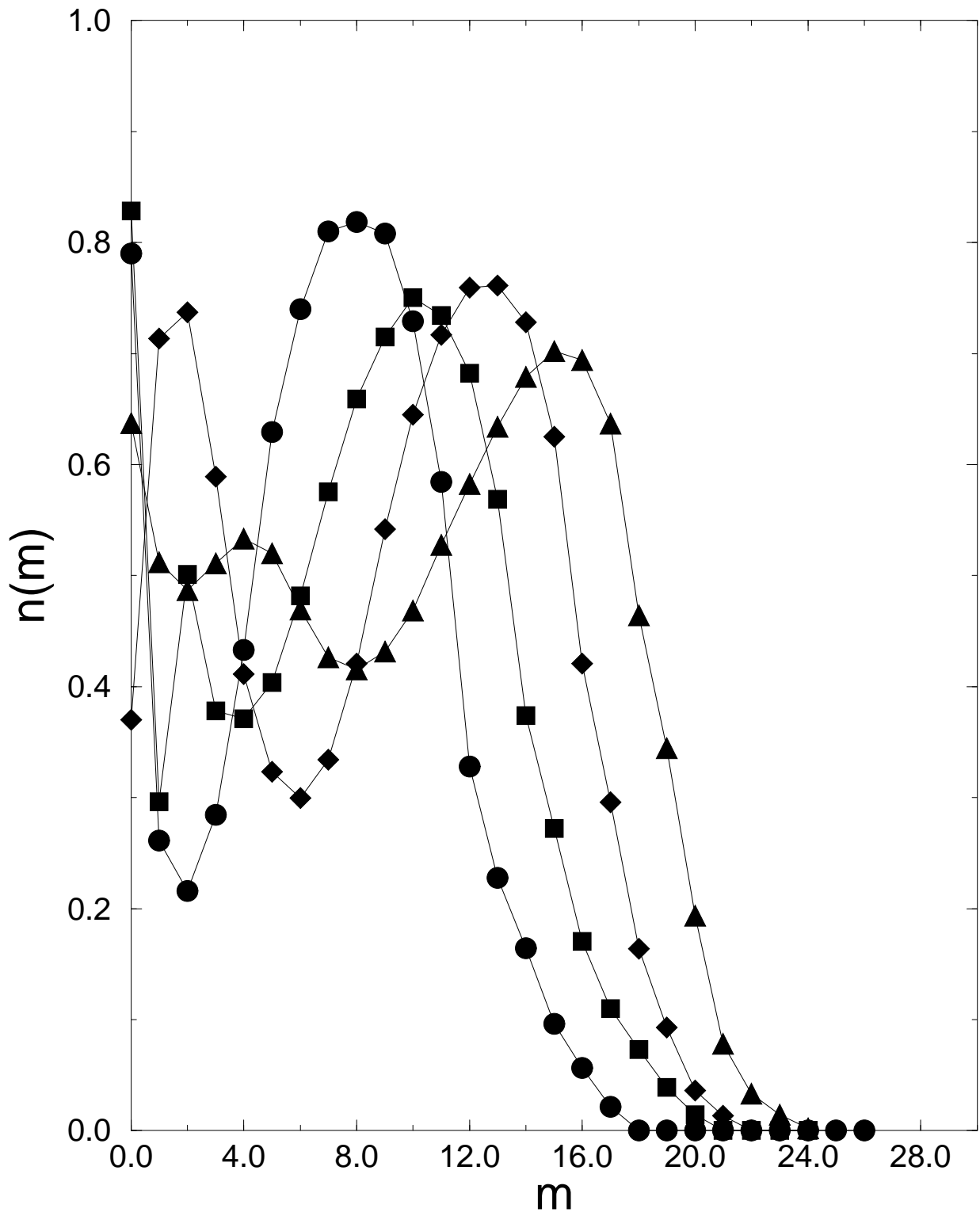
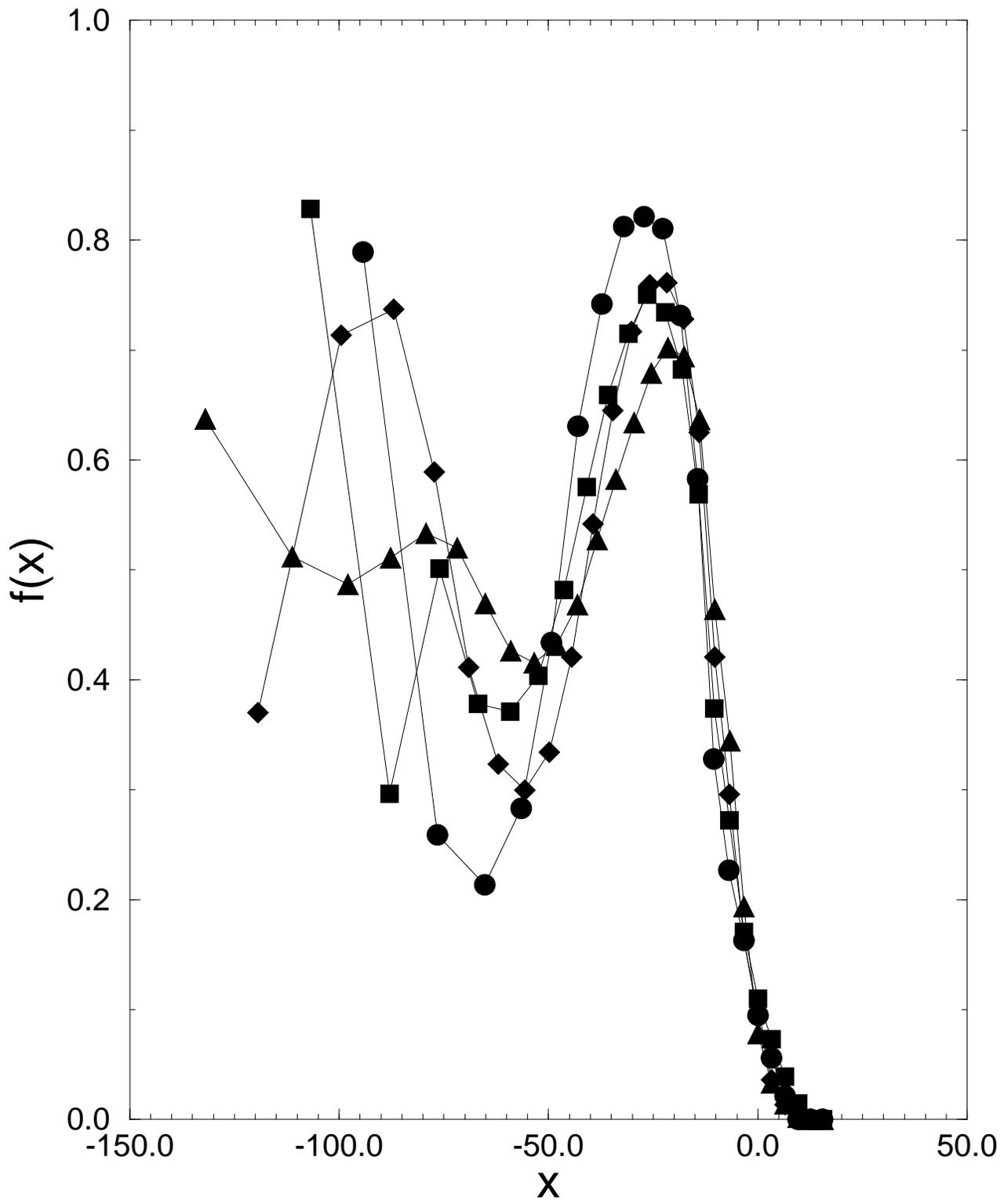
\end{document}